\documentclass{article}

\usepackage{arxiv}

\usepackage{doc}
\usepackage{makeidx}
\usepackage{amssymb}
\usepackage{graphicx,times,amsmath,cite}
\usepackage{epstopdf}
\usepackage{url}
\usepackage{subfig}
\usepackage{comment} 
\usepackage{mathrsfs} 
\usepackage{multirow}
\usepackage{rotating}
\usepackage{array}
\usepackage{makecell}
\usepackage{comment} 
\usepackage{algorithm}
\usepackage{algorithmic}

\title{Pattern Denoising in Molecular Associative Memory \\ using Pairwise Markov Random Field Models}
\author{Dharani Punithan \\punithan.dharani@gmail.com\\ Institute of Computer Technology \\Seoul National University\\ South Korea
	\And
	Byoung-Tak Zhang \\btzhang@snu.ac.kr \\Dept. of Computer Science and Engineering\\ Seoul National University\\ South Korea 
}

\date{}

\renewcommand{\headeright}{Pattern Denoising in Molecular Associative Memory}
\renewcommand{\undertitle}{}

\begin{document}

\maketitle


\begin{abstract}
We propose an \textit{in silico} molecular associative memory model for pattern learning, storage and denoising using Pairwise Markov Random Field (PMRF) model. Our PMRF-based molecular associative memory model extracts locally distributed features from the exposed examples, learns and stores the patterns in the molecular associative memory and denoises the given noisy patterns via DNA computation based operations. Thus, our computational molecular model demonstrates the functionalities of content-addressability of human memory. Our molecular simulation results show that the averaged mean squared error between the learned and denoised patterns are low ($<0.014$) up to $30$\% of noise. 
\end{abstract}	

\section{Introduction}
Memory is a crucial part of the learning process in both animals and humans. It is the mental process of encoding, storing and retrieving. Among the different types of memories, the most useful one is associative memory (AM). AM stores data in a distributed fashion and is addressed through its contents. Hence, AM is also known as a content-addressable memory (CAM)~\cite{Kohonen89a}. AM works by learning patterns and retrieving or reconstructing a previously learned pattern that most closely resembles the noisy patterns. Thus, it has applications in pattern matching, pattern recognition, robotics, etc. 
This type of memory is robust and fault-tolerant as it exhibits error-correction ability. 

DNA works as a ``memory'' to store genetic information in the cellular organism. The striking features of DNA~\cite{Baum583, Adleman1021, rothemund2004algorithmic} such as self-assembly, huge information storage capacity and massive parallelism are similar to the brain~\cite{chklovskii2004cortical, zaidel2003parallel, whitesides2002self}. Hence, associative memory can be realized on a molecular level, which can be vaster than human brain~\cite{Baum583, reif2001experimental}. Some recent studies~\cite{qian2011neural,macvittie2013bioinspired} show that  molecular systems can exhibit brain-like cognition. In our \textit{in silico} molecular simulation, we demonstrate the potential of molecular associative memory using the popular image processing tool PMRF. To our knowledge, our work is the first to denoise patterns using molecular algorithms.

This work is based on~\cite{zhang2008hypernetworks} and also an extension of our previous works~\cite{punithan2018molecular, punithan2017ising}. In our previous works, we used mutation in the learning phase, which we avoid in this work so as to match the conventional DNA computing based bio-algorithms. In addition to the recall functionality as in our previous works, here we propose molecular methods to denoise the noisy patterns iteratively using the PMRF model.
To summarize, the tasks of our proposed molecular associative memory model are 1) to learn and store a set of patterns (digits from $0$ to $9$) when exposed to MNIST~\cite{lecun1998mnist} training examples and 2) to denoise the noisy patterns iteratively. We combine DNA-based bio-molecular operations such as hybridization, melting, and amplification with PMRF model to demonstrate these functionalities. We use PMRF formulations, but the involved computations are based on hybridization reactions. We mainly take advantage of the hybridization operations to implement the proposed molecular content-addressable memory.

\section{Background}
\subsection{Pairwise Markov Random Field Models}
\label{bg:mrf}
Consider an undirected graph $G=(\mathcal{V},E)$ on a two-dimensional lattice $L$, where nodes ($\mathcal{V}$) represent random variables $\{X_{ij}\}$ and edges ($E$) represent (conditional dependencies) association between two nodes. Such a graph is called a Markov Random Field (MRF)~\cite{li2012markov, won2013stochastic, beckerman1997adaptive}, $X$, if it holds the Markovian property $p(x_{ij}|x_{kl}, (k,l) \in L, (k,l) \neq (i,j)) = p(x_{ij}|x_{kl}, (k,l) \in \mathcal{N}_{ij}); \forall (i,j) \in L$, where $i$ and $k$ are row indices, $j$ and $l$ are column indices, $x_{ij}$ and $x_{kl}$ are the realizations of the random variables associated with the specified lattice points and $\mathcal{N}_{ij}$ is the neighborhood of $(i,j)$. The local conditional probability for each node can be defined using the clique potentials. A clique ($c$) is defined to be either a single node or a collection of nodes in which every node is a neighbor of every other node. Each clique is specified a potential $V_{c}(.)$. The sum of all clique potentials for a realization $x$ is called the energy function ~\cite{li2012markov} : 

\begin{equation}
\label{eq:energy} 
U(x) = \sum_{ij \in L}  \sum_{c \in \mathcal{C}_{ij}} V_{c}(x)
\end{equation}
where $\mathcal{C}_{ij}$ is the set of all cliques associated with node $x_{ij}$ and $V_{c}(x)$ is the clique potential associated with a clique $c$.

The number of nodes in a clique is called the order of the clique.
Potentials of order one and two are called unary and pairwise respectively. A Pairwise Markov Random Field (PMRF)~\cite{li2012markov, won2013stochastic, beckerman1997adaptive} over the graph is associated with a set of unary (node) potentials and a set of pairwise (edge) clique potentials; which implies the order of clique size is (at most) two. 
For PMRF, the energy is defined as~\cite{li2012markov} :

\begin{equation}
\label{eq:pEnergy} 
U(x) = \sum_{ij \in L} \Bigg[ V_{1}(x_{ij}) + \sum_{ kl \in \mathcal{N}_{ij}}  V_{2}(x_{ij}, x_{kl}) \Bigg]
\end{equation}

where $V_{1}(x_{ij})$ and $V_{2}(x_{ij},x_{kl})$ are the unary and the pairwise clique potentials respectively.
The local conditional probabilities for PMRF are defined as in the equation~\ref{eq:condProb} ~\cite{li2012markov}. 

\begin{equation}
\label{eq:condProb}
p(x_{ij} | x_{kl}, kl \in \mathcal{N}_{ij}) = 
\frac{exp \Big[V_{1}(x_{ij}) + \sum_{ kl \in \mathcal{N}_{ij}}  V_{2}(x_{ij}, x_{kl})  \Big] }{\sum_{x_{ij}} exp \Big[V_{1}(x_{ij}) + \sum_{ kl \in \mathcal{N}_{ij}}  V_{2}(x_{ij}, x_{kl})  \Big]}
\end{equation}

PMRFs are attractive because of their simplicity. These graphical models are popular in the field of statistical physics and have applications in computer vision, computational biology, information extraction, etc.

\subsection{DNA based Bio-molecular Operations}
DNA consists of four different bases A (Adenine), T (Thymine), C (Cytosine) and G (Guanine). These bases are connected together to form a single-stranded DNA sequence. Two single strands bind to form a double-stranded DNA helix by Watson-Crick complementary rule~\cite{watson1953dna} whereby adenine bonds with thymine (A-T) and vice versa (T-A), cytosine bonds with guanine (C-G) and vice versa (G-C). This base-pairing of complementary single-stranded molecules to form a double-stranded DNA is called \textit{hybridization} (or annealing). The reverse process, a double-stranded helix yielding its two constituent single-strands, is called \textit{melting} (or denaturation). The process of multiplying the copies of DNA strands is called \textit{amplification}. 

\section{Methods}
\subsection{Molecular Memory and Encoding}

Molecular memory is modeled as a set of $m$ two-dimensional weighted graphs $M=\{G^{m}=(\mathcal{V}^{m}, \mathcal{C}^{m}, W^{m})\}$, each of size $N \times N$, where $m$ represents the number of binary patterns to be learned (digits from $0$ to $9$), $\mathcal{V}^{m}$ is a set of all nodes representing pixels $\{x_{ij}^{m}\}$ of the $m\textsuperscript{th}$ pattern, $i$ and $j$ represent row and column indices of the pixel location, $\mathcal{C}^{m}$ is the set of all unary (first-order) and pairwise (second-order) cliques 
in the second-order (8-point) neighborhood system of the $m\textsuperscript{th}$ pattern and $W^{m}$ represents the weights of the nodes of the $m\textsuperscript{th}$ pattern. We set $N=28$; as each MNIST example is of size $28 \times 28$. 

In our previous works~\cite{punithan2018molecular, punithan2017ising}, all pixels of all patterns in the memory were initially black. On training, we extracted the information from the training examples and the molecular memories were mutated with respect to the foreground pixels of the training images. In this work, we avoid mutation to match the DNA computing based bio-algorithms. Hence, we create all possible unary and pairwise cliques in the initial memory. For each pixel location (row and column indices), we create both black (background pixel) and white (foreground pixel) DNA molecules, as we learn binary patterns. Then, for each pixel location and pixel color, we create all possible unary and pairwise cliques. We construct $m$ such bags of DNA single-strands. Each  single-strand represents either a unary (pixel) or a pairwise clique. We form the molecules from the four-letter DNA alphabet ${A, T, G, C}$. For example, a pixel (node) information -- location (row and column indices) and color (black or white) -- is encoded into a DNA sequence as `GTGGTTA'; `GTG' (first three bases) represent row index ($i$) of a pixel, `GTT' (next three bases) represent column index ($j$) of that pixel and `A' (last base) represents the color of that binary pixel. We combine two such sequences to form the pairwise cliques. 

We then re-encode the character-based DNA sequence into a $2 \times n$ matrix, where $n$ is the number of bases of the DNA sequence. Each DNA base is re-encoded into a vector : $A$ as $[1, 0]^{T}$, $T$ as $[-1, 0]^{T}$, $G$ as $[0, 1]^{T}$, and $C$ as $[0, -1]^{T}$. This initial molecular memory is trained on the MNIST examples to memorize patterns (digits from $0$ to $9$). During learning, the weights of the unary single-strands of the memory are updated using the conditional probabilities (refer algorithm~\ref{algoLearning}). After memorization, the model recalls a stored pattern, which has the maximum weighted score of the DNA molecules, to the given noisy pattern. We then denoise the given noisy pattern iteratively by computing energies (refer algorithm~\ref{algoDenoise}).  We use PMRF model for computing conditional probabilities, weighted scores and energies. These formulations depend only on clique potentials. We can define our own clique potentials according to our problem, as long as they  emphasize some specific features~\cite{won2013stochastic}. In our modeling, we define clique potentials in terms of hybridization reactions. On a complete hybridization, we assign $1$ to the respective clique potential; or else $0$.

\subsection{Molecular Learning and Storage (Memorization)}
\begin{algorithm} [H]	
	Input: Initial molecular memory and MNIST training examples.\\
	Output: A set of learned molecular patterns from $0$ to $9$ ($M$).
	\caption{Molecular Learning and Storage of Patterns}
	\label{algoLearning}
	\begin{itemize}	
		\item Loop over MNIST training examples
		\subitem Read the grayscale MNIST image and get the label 
		\subitem Binarize the MNIST image and remove noise
		\subitem Encode each pixel information of the image into character-based DNA molecules
		\subitem Form single-strands of unary and pairwise cliques of DNA molecules    
		\subitem Re-encode character-based DNA molecules into vector-based numerical DNA molecules 	
		\subitem \textbf{Hybridization:} Bind single-strands of MNIST image with  single-strands of $m\textsuperscript{th}$ memory pattern matching MNIST label			
		\begin{algorithmic}[t]
			\IF{(hybridization reactions are complete)}
			\STATE the corresponding memory clique potentials ($V_{1}(x_{ij}^{m})$ and $V_{2}(x_{ij}^{m}, x_{kl}^{m})$)  are set to $1$
			\ELSE
			\STATE clique potentials are set to $0$
			\ENDIF
		\end{algorithmic}
		\subitem \textbf{Melting:} Separate single-strands of MNIST image and  $m\textsuperscript{th}$ memory pattern				
		\subitem Compute local conditional probabilities $p(x_{ij}^{m}=1 | x_{\mathcal{N}_{ij}}^{m})$ with clique potentials
		\subitem \textbf{Amplification:} Update the weights ($w_{ij}^{m}$) of single-strands of the memory
		using \\conditional probabilities	
		\item Set weights of the memory single-strands lesser than $0.002$ to $0$
		\item Normalize the weights of the memory single-strands
	\end{itemize}
\end{algorithm}

Each MNIST ($G^{m^{t}}$) image is mapped to a realization of a PMRF such that nodes represent pixels ($x_{ij}^{m^{t}}$) of the image. We binarize each grayscale MNIST training image and remove noise if present. Each pixel in the image comprises the location (row and column) and the color (black or white) information. Each pixel information is encoded into character-based DNA molecules. The DNA molecules, representing the pixel locations, are complementary to memory strands. We form unary and pairwise single-strands in the second-order neighborhood system. We re-encode each character-based DNA molecules into respective vector-based numerical DNA molecules, as mentioned before. The single-strands of the training image are hybridized with the single-strands of $m\textsuperscript{th}$ memory pattern corresponding to the label of that training image. The addition of two single-strands (one from memory and another from the training example) yielding a zero matrix, indicates a complete hybridization. On complete hybridizations, the clique potentials ($V_{1}(x_{ij}^{m})$ and $V_{2}(x_{ij}^{m}, x_{kl}^{m})$) are set to one; otherwise zero. We separate the set of DNA single-strands, representing the training example, exposed in that iteration; this is known as melting operation. Then, we compute the conditional probabilities of the foreground (white) pixels ($x_{ij}^{m}=1$) given the neighborhood ($x_{\mathcal{N}_{ij}}^{m}$) at the pixel location $(i,j)$ of the $m\textsuperscript{th}$ memory pattern based on equation~\ref{eq:condProb}. 
For the binary random variables, the conditional probabilities     are computed as in equation~\ref{eq:condProbability}~\cite{beckerman1997adaptive}.

\begin{equation}
\label{eq:condProbability}
p(x_{ij}^{m}=1 | x_{\mathcal{N}_{ij}}^{m}) = 
\frac{exp \Big[V_{1}(x_{ij}^{m}) + \sum_{ kl \in \mathcal{N}_{ij}}  V_{2}(x_{ij}^{m}, x_{kl}^{m})  \Big]}{1 + exp \Big[V_{1}(x_{ij}^{m}) + \sum_{ kl \in \mathcal{N}_{ij}}  V_{2}(x_{ij}^{m}, x_{kl}^{m})  \Big]}
\end{equation}

We use the computed conditional probabilities (refer equation~\ref{eq:condProbability}) to update the weights ($w_{ij}^{m}$) of the foreground pixels of the $m\textsuperscript{th}$ memory pattern (refer equation~\ref{eq:weightUpdate}).
\begin{equation}
\label{eq:weightUpdate} 
w_{ij}^{m}(new) = w_{ij}^{m}(old) + \eta * p(x_{ij}^{m}=1 | x_{\mathcal{N}_{ij}}^{m})
\end{equation}
where $\eta=1/(1+exp(-\gamma *(iterNum-stepSize)))$ is sigmoid decay learning rate, $\gamma=0.01$ is decay rate, iterNum is the current training iteration number and stepSize $=100$. This weight update step is referred to as amplification.
Learning strengthens the weight of respective foreground pixels of the memory. After training, we set the weights of the foreground pixel, having smaller weights ($<0.002$), to zero. Then, we normalize the weights so that the sum of all the weights of the foreground pixels is one. This whole learning procedure is given in the algorithm~\ref{algoLearning}.

\subsection{Molecular Denoising of Noisy Patterns}

\begin{algorithm}[!htb] 
	Input: Noisy stored pattern and a set of memory patterns ($M$) (from $0$ to $9$).
	\\ Output: Denoised pattern.
	\caption{Molecular Denoising of Noisy Patterns}
	\label{algoDenoise}	
	\begin{itemize}					
		\item Read the given noisy pattern 
		\item Encode each pixel into character-based DNA molecules	
		\item Form single-strands of unary and pairwise cliques of DNA molecules	
		\item Re-encode each character-based DNA molecule into numerical DNA molecules
		\item Get best-matched pattern from memory  	
		\begin{algorithmic}[t]
			\FOR{$m=0$ to $9$} 
			\item \textbf{Hybridization:} Bind single-strands of noisy image and $m\textsuperscript{th}$ memory pattern 	
			\item Set memory potentials ($V_{1}(x_{ij}^{m})$ and $V_{2}(x_{ij}^{m},x_{kl}^{m})$) to $1$ on complete hybridizations						
			\item Compute weighted score ($score^m$) for the $m\textsuperscript{th}$ memory pattern
			\item \textbf{Melting:} Separate the  single-strands of noisy image	and $m\textsuperscript{th}$ memory pattern
			\item $m := m+1$
			\ENDFOR
			\\ Compute softmax $\sigma(score^m)$ of the scores and get the label of the best-matched memory pattern
		\end{algorithmic}	
		
			\item \begin{algorithmic}
				\IF{(label of best-matched memory pattern $==$ label of noisy pattern)}					
				\STATE{\textbf{Hybridization:} Bind single-strands of noisy image and $m\textsuperscript{th}$ memory pattern}
				\STATE Set noisy potentials ($V_{1}(x_{ij}^{m^\prime})$ and $V_{2}(x_{ij}^{m^\prime},x_{kl}^{m^\prime})$)	to $1$ on complete hybridizations					
				\STATE{oldEnergy := Compute the energy of the noisy pattern based on hybridization reactions}						
				\STATE	Loop over the pixels of the noisy pattern
					\subitem{Pick a random pixel from the noisy pattern}				
						\IF{(the pixel color is black)}
						\STATE Add new character-based DNA molecules for that location with pixel color white 
						\ELSIF{(the pixel color is white)}
						\STATE Add new character-based DNA molecules for that location with pixel color black 							
						\ENDIF				
					\subitem Add all possible unary and pairwise cliques correspondingly
					\subitem Re-encode the new character-based DNA molecules to numerical DNA molecules
					\subitem{\textbf{Hybridization:} Bind the new noisy single-strands with memory single-strands}	
					\subitem Set noisy potentials ($V_{1}(x_{ij}^{m^\prime})$ and $V_{2}(x_{ij}^{m^\prime},x_{kl}^{m^\prime})$) to $1$ on complete hybridizations									
					\subitem{newEnergy := Compute the energy with the changed pixel color}

					\IF{($newEnergy < oldEnergy$)}					 
						\STATE {oldEnergy := newEnergy}
					    \STATE \textbf{Melting} : Separate old single-strands
				    \ELSE
					     \STATE \textbf{Melting} : Separate newly added single-strands
					\ENDIF			
					\ENDIF	
				
			\end{algorithmic}				
		\end{itemize}
\end{algorithm}

Our model finds the closest memory pattern to a given noisy pattern ($G^{m^{\prime}}$) by applying hybridization operations and computing the weighted average of the local clique potentials. Each pixel ($x_{ij}^{m^\prime}$) information  in the noisy pattern is encoded into character-based DNA molecules. The encoded location information is complementary to the memory DNA molecules representing the location. The unary and pairwise cliques of DNA sequences at each pixel location in second-order neighborhood system of the noisy pattern are formed. We then re-encode each character-based DNA molecule into numerical vector-based DNA molecules. The single-strands of the noisy pattern are hybridized with single-strands of the memory. 
The clique potentials ($V_{1}(x_{ij}^{m})$ and $V_{2}(x_{ij}^{m}, x_{kl}^{m})$) are set to one on complete hybridizations; otherwise to zero. The weighted score (refer equation~\ref{eq:score}) is computed for each of the memory pattern (digits from $0$ to $9$) and the softmax (refer equation~\ref{eq:softmax}) of scores is computed to retrieve the closest memory pattern. All single-strands of the noisy pattern are separated after each weighted score computation. 

\begin{equation}
\label{eq:score}
score^m = \sum_{x_{ij}^{m}=1} w_{ij}^{m} * \Bigg[ \frac{V_{1}(x_{ij}^{m})+ \sum\limits_{kl \in \mathcal{N}_{ij}} V_{2}(x_{ij}^{m}, x_{kl}^{m})}{|\mathcal{C}_{ij}|} \Bigg]; \quad m=0,...,9.
\end{equation}

where $|\mathcal{C}_{ij}|$ is the cardinality of the clique set.

\begin{equation}
\label{eq:softmax}
\sigma(score^m)= \frac{\exp(score^m)}{\sum\limits_{l=0}^{9} \exp(score^l)}; \quad m=0,...,9.
\end{equation}

The next step is to denoise the noisy pattern. We hybridize the single-strands of the best-matched memory pattern with the single-strands of the noisy pattern. On complete hybridizations, the clique potentials ($V_{1}(x_{ij}^{m^\prime})$ and $V_{2}(x_{ij}^{m^\prime}, x_{kl}^{m^\prime})$) are set to one; otherwise zero. We then compute the energy as defined in the equation~\ref{eq:netEnergy}. 

\begin{equation}
\label{eq:netEnergy} 
U(x^{m^\prime}) = \sum_{ij}  \Bigg[ V_{1}(x_{ij}^{m^\prime}) +  \sum_{ kl \in \mathcal{N}_{ij}}  V_{2}(x_{ij}^{m^\prime}, x_{kl}^{m^\prime})\Bigg]   
\end{equation}

We randomly pick a pixel location and if the color of the pixel is black (white), we add the new character-based DNA molecules with pixel color white (black) for that location. We form possible pairwise cliques. We then re-encode them into numerical DNA molecules. We hybridize the newly added single-strands with the memory single-strands and we compute the energy again (refer equation~\ref{eq:netEnergy}). The DNA molecules corresponding to the higher energy are separated by melting. We repeat this process for all the pixels in the noisy pattern.

\section{Results}
\label{sec:result}

In this section, we present the results of tasks of our molecular content-addressable memory model : 1) learning of patterns from exposed examples and storage in memory and 2) denoising of given noisy patterns.

\subsection{Learned and Stored Patterns in Molecular Associative Memory}
\begin{figure}[!htb] 	
	\centering
	\subfloat{
		\label{zero}
		\includegraphics[width=2 cm]{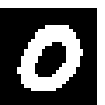}
	}
	\subfloat{
		\label{one}
		\includegraphics[width=2 cm]{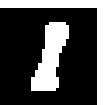}
	}
	\subfloat{
		\label{two}
		\includegraphics[width=2 cm]{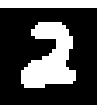}
	}
	\subfloat{
		\label{three}
		\includegraphics[width=2 cm]{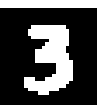}
	}
	\subfloat{
		\label{four}
		\includegraphics[width=2 cm]{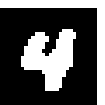}
	}
	\\
	\subfloat{
		\label{five}
		\includegraphics[width=2 cm]{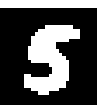}
	}
	\subfloat{
		\label{six}
		\includegraphics[width=2 cm]{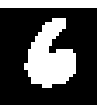}
	}
	\subfloat{
		\label{seven}
		\includegraphics[width=2 cm]{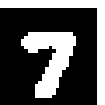}
	}
	\subfloat{
		\label{eight}
		\includegraphics[width=2 cm]{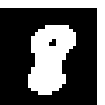}
	}
	\subfloat{
		\label{nine}
		\includegraphics[width=2 cm]{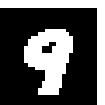}
	}
	\caption{Patterns Learned and Stored in Molecular Associative Memory} 
	\label{fig:storedPatterns}
\end{figure}

Figure~\ref{fig:storedPatterns} shows the learned and stored patterns in our molecular associative memory. We use MNIST training dataset to train the model. We use $50,000$ training examples; $5000$ for each digit (from $0$ to $9$). We extract the features (unary and pairwise cliques) from these training examples, encode them to DNA molecules and learn the exposed examples via DNA computing based bio-operations. On training, the weights of the foreground pixels of the memory patterns are strengthened by computing PMRF  based conditional probabilities. In our modeling, local conditional probabilities are computed in terms of hybridization reactions.

\subsection{Noisy Dataset}
\begin{figure}[!htb] 	
	\centering
	\subfloat[Noise$=10\%$]{
		\label{noise10}
		\includegraphics[width=2 cm]{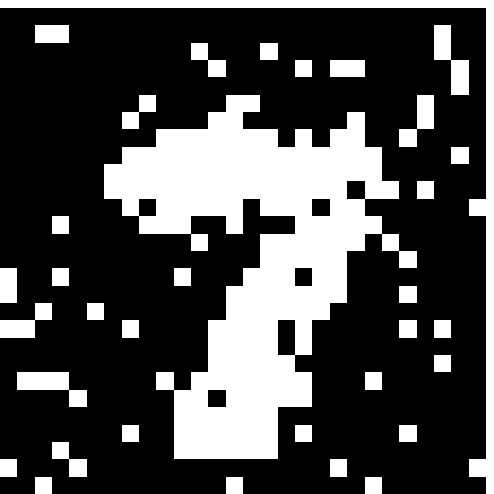}
	}
	\subfloat[Noise$=20\%$]{
		\label{noise20}
		\includegraphics[width=2 cm]{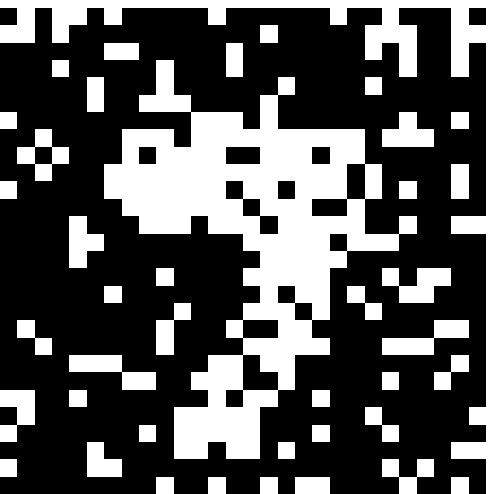}
	}
	\subfloat[Noise$=30\%$]{
		\label{noise30}
		\includegraphics[width=2 cm]{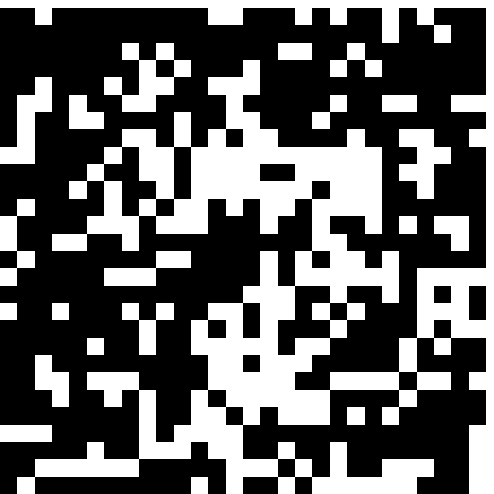}
	}
	\subfloat[Noise$=40\%$]{
		\label{noise40}
		\includegraphics[width=2 cm]{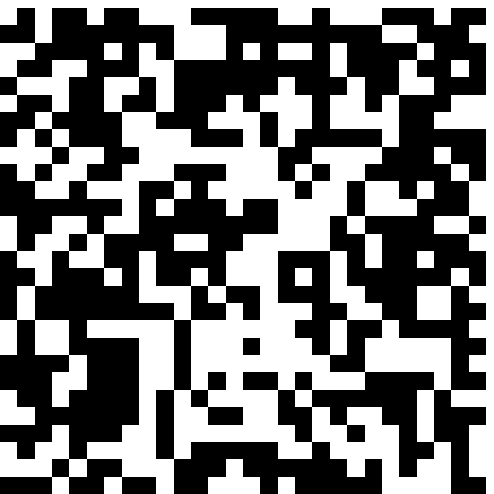}
	}		
	\subfloat[Noise$=50\%$]{
		\label{noise50}
		\includegraphics[width=2 cm]{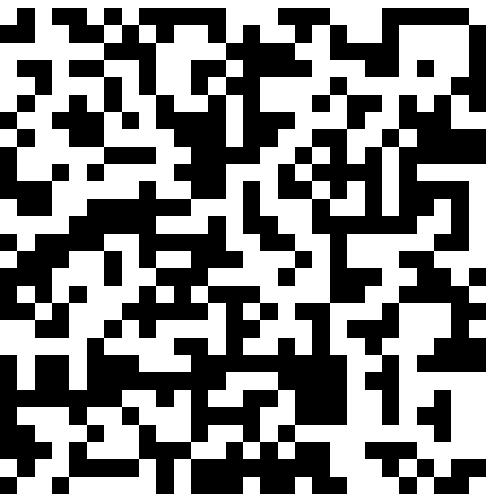}
	}		
	\caption{Noisy Versions of pattern $7$ at Different Random Noise Percentages ($\%$)}
	\label{fig:noisyPatterns}
\end{figure}

We create an artificial dataset by adding random noise to the learned patterns at different noise percentages (from $0\%$ to $50\%$). We randomly change the pixel color (from black to white and vice versa) of the patterns to the given noise percentage. We then encode each pixel to DNA molecules for further processing.
Adding $50\%$ of noise makes random patterns. Hence, we examine our model up to $50\%$ of noise. For each pattern and for each noise percentage, we create $100$ noisy patterns and hence $6,000$ in total. The noisy samples of pattern $7$ at different noise percentages 
are shown in  figure~\ref{fig:noisyPatterns}. 

\subsection{Denoised Patterns}

\begin{figure}[!htb]
	\includegraphics[angle=-90, width=9cm]{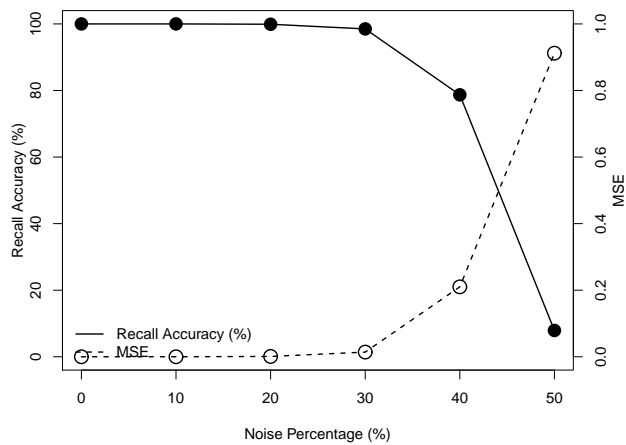}
	\centering
	\caption{Recall Accuracy (\%) and MSE Vs. Noise Percentage (\%)} 
	\label{fig:accVsNoise}
\end{figure}

\begin{figure}[!htb] 	
	\centering
	\subfloat{
		\label{0_0}
		\includegraphics[width=2 cm]{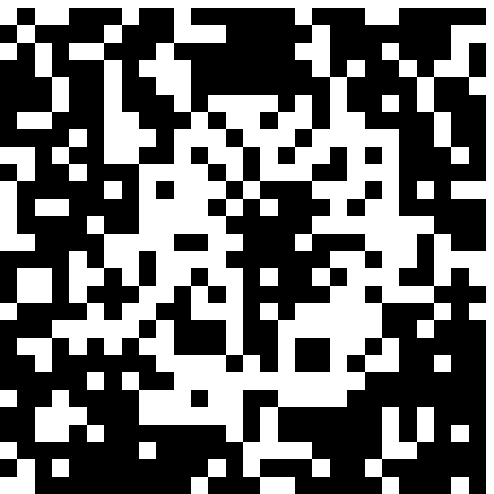}
	}
	\subfloat{
		\label{0_300}
		\includegraphics[width=2 cm]{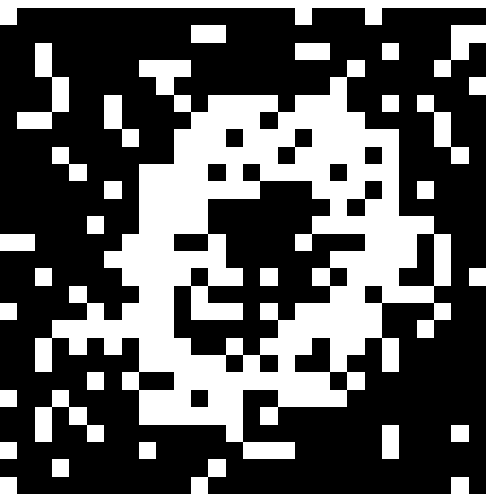}
	}
	\subfloat{
		\label{0_500}
		\includegraphics[width=2 cm]{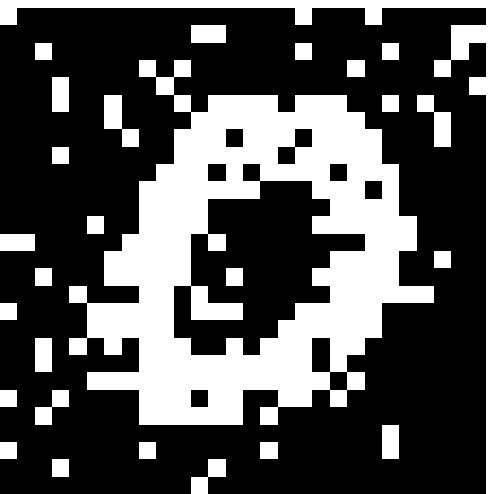}
	}
	\subfloat{
		\label{0_600}
		\includegraphics[width=2 cm]{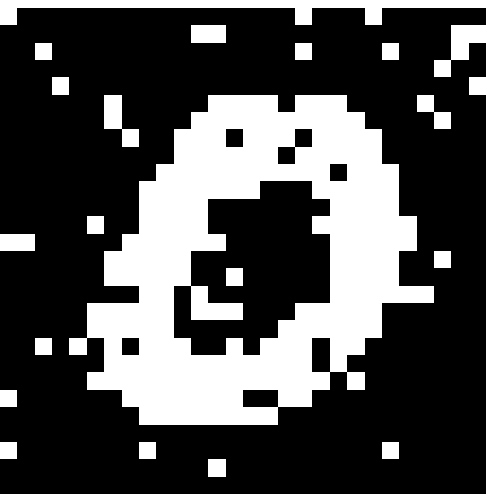}
	}
	\subfloat{
		\label{0_784}
		\includegraphics[width=2 cm]{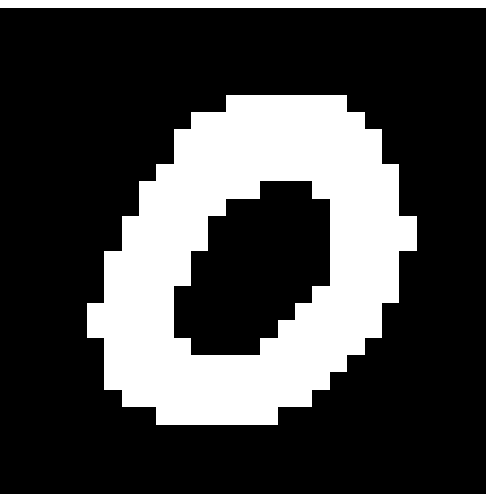}
	}    
	\\
	\subfloat{
		\label{1_0}
		\includegraphics[width=2 cm]{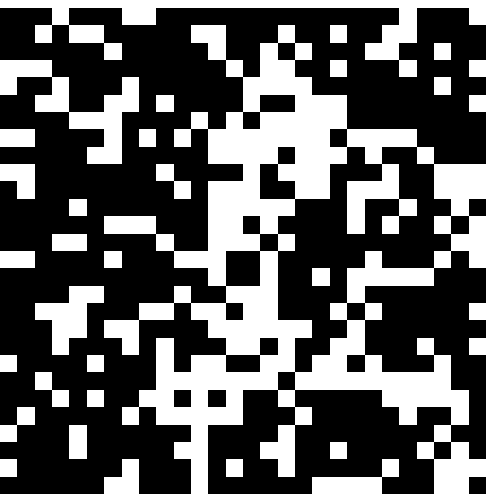}
	}
	\subfloat{
		\label{1_300}
		\includegraphics[width=2 cm]{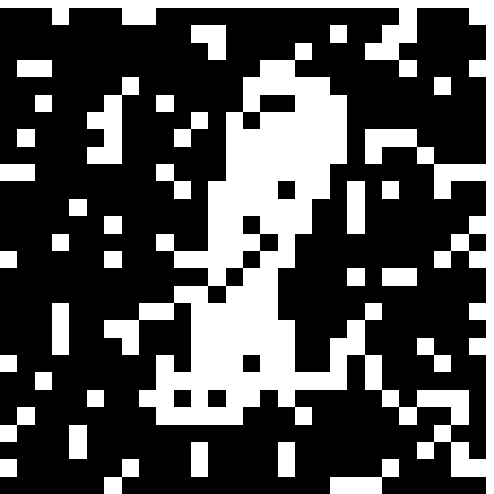}
	}
	\subfloat{
		\label{1_500}
		\includegraphics[width=2 cm]{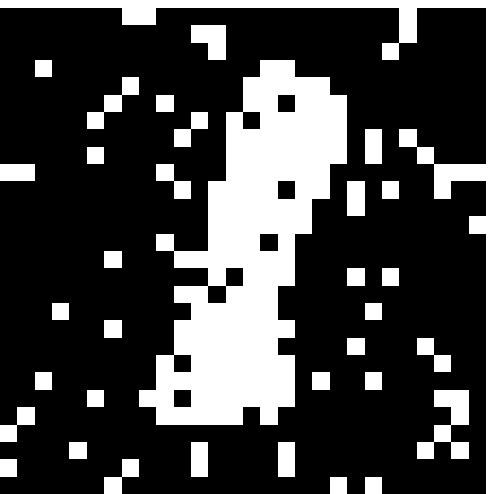}
	}
	\subfloat{
		\label{1_600}
		\includegraphics[width=2 cm]{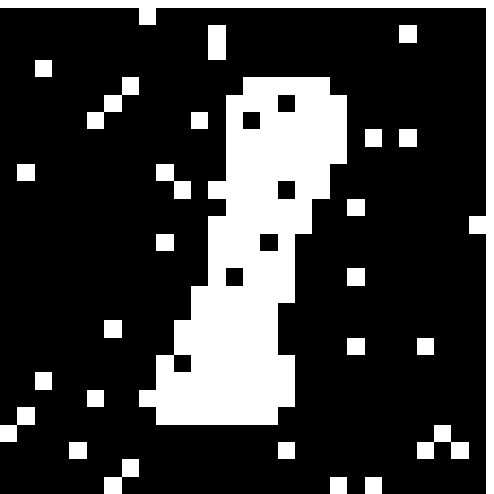}
	}
	\subfloat{
		\label{1_784}
		\includegraphics[width=2 cm]{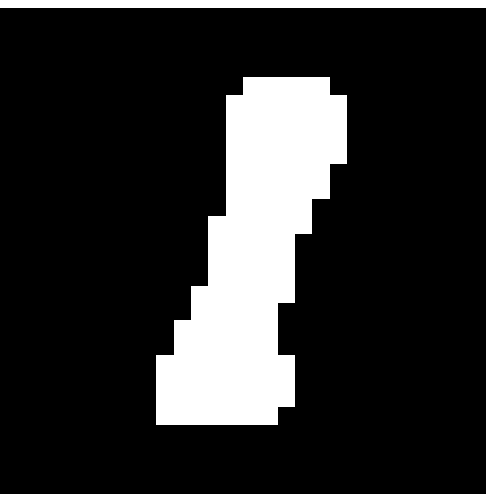}
	}
	\\
	\subfloat{
		\label{2_0}
		\includegraphics[width=2 cm]{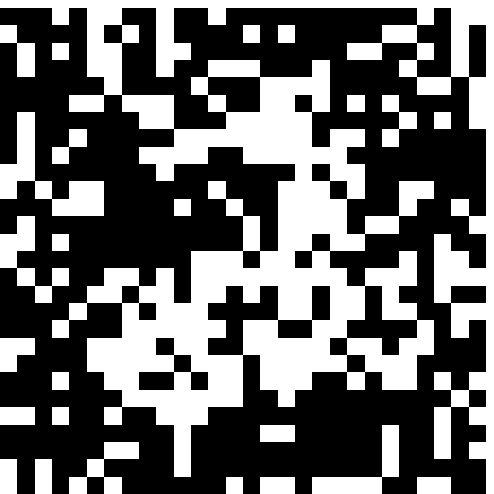}
	}
	\subfloat{
		\label{2_300}
		\includegraphics[width=2 cm]{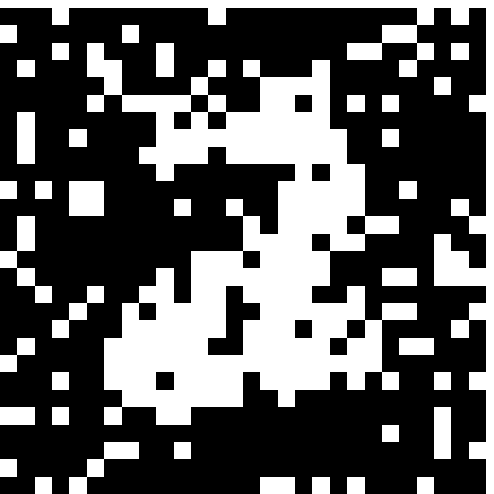}
	}
	\subfloat{
		\label{2_500}
		\includegraphics[width=2 cm]{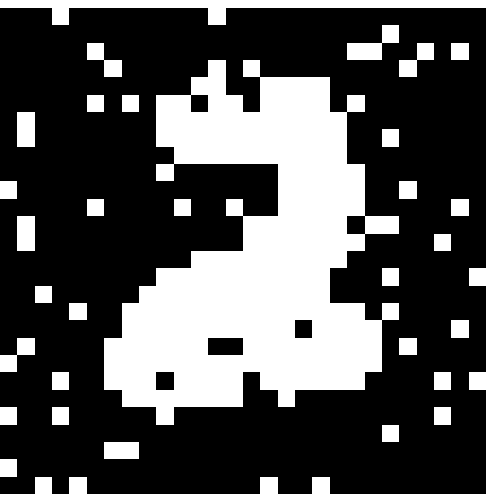}
	}
	\subfloat{
		\label{2_600}
		\includegraphics[width=2 cm]{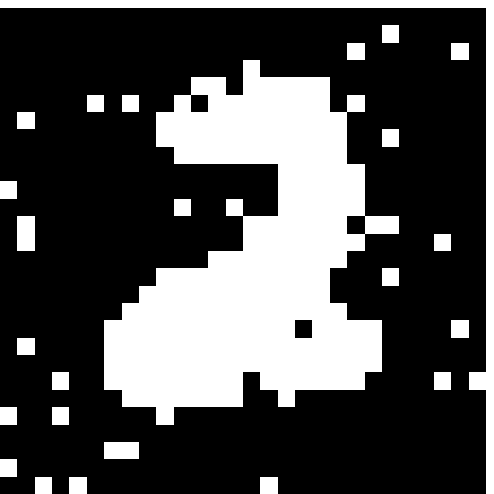}
	}
	\subfloat{
		\label{2_784}
		\includegraphics[width=2 cm]{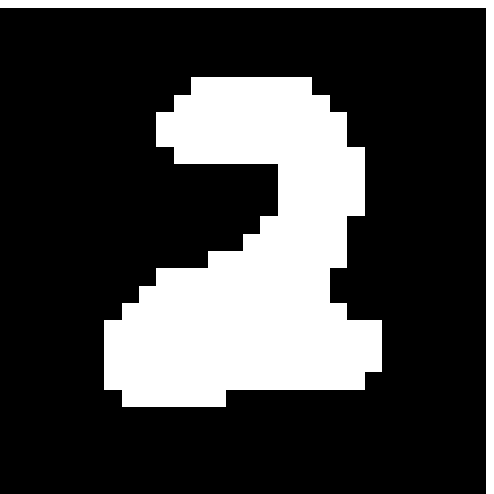}
	}
	\\
	\subfloat{
		\label{3_0}
		\includegraphics[width=2 cm]{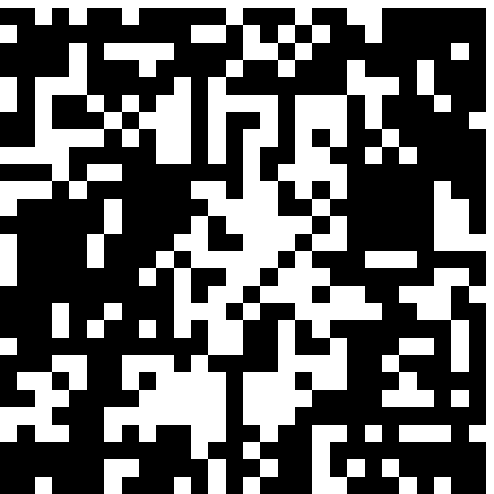}
	}
	\subfloat{
		\label{3_300}
		\includegraphics[width=2 cm]{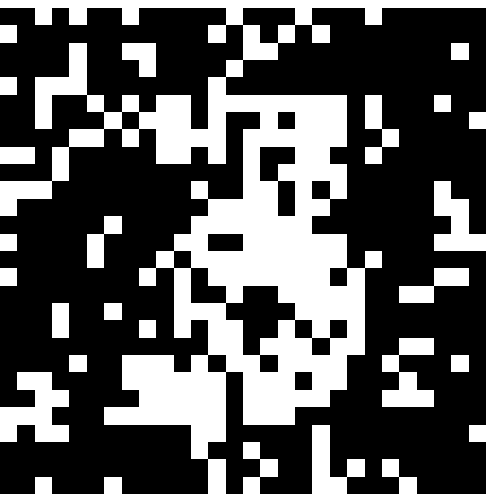}
	}
	\subfloat{
		\label{3_500}
		\includegraphics[width=2 cm]{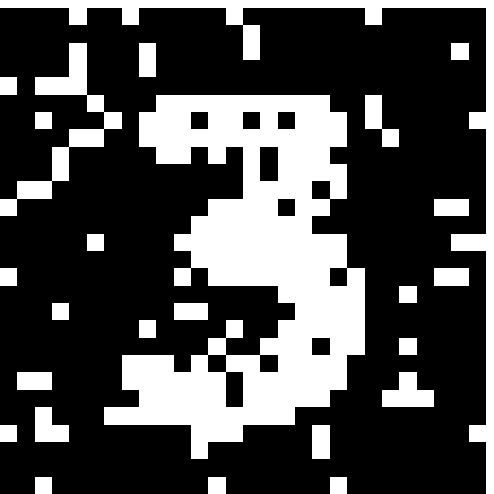}
	}
	\subfloat{
		\label{3_600}
		\includegraphics[width=2 cm]{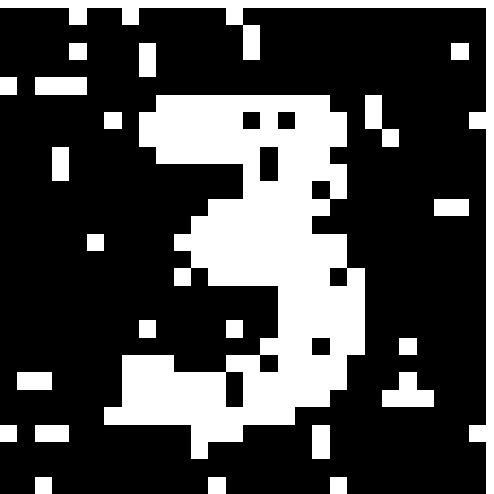}
	}
	\subfloat{
		\label{3_784}
		\includegraphics[width=2 cm]{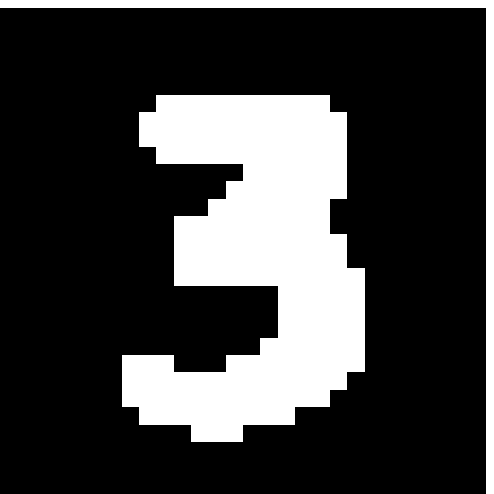}
	}
	\\
	\subfloat{
		\label{4_0}
		\includegraphics[width=2 cm]{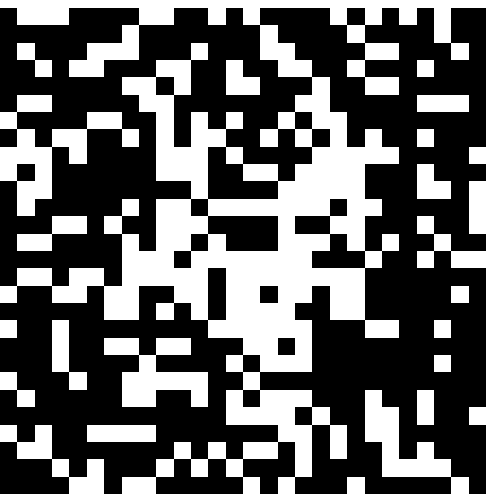}
	}
	\subfloat{
		\label{4_300}
		\includegraphics[width=2 cm]{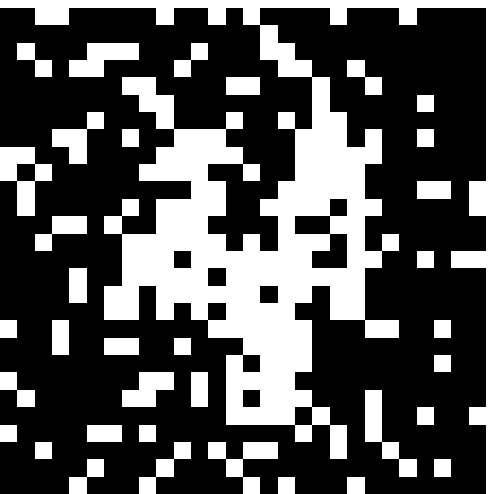}
	}
	\subfloat{
		\label{4_500}
		\includegraphics[width=2 cm]{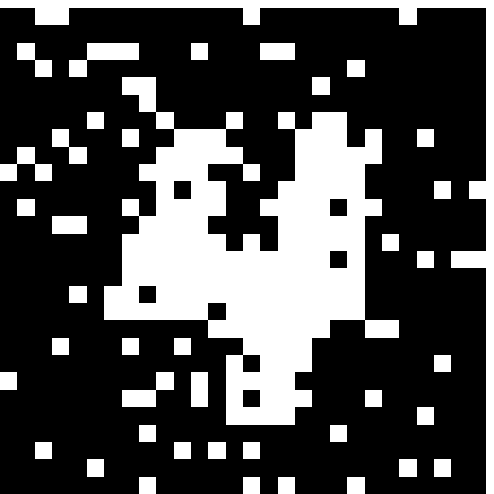}
	}
	\subfloat{
		\label{4_600}
		\includegraphics[width=2 cm]{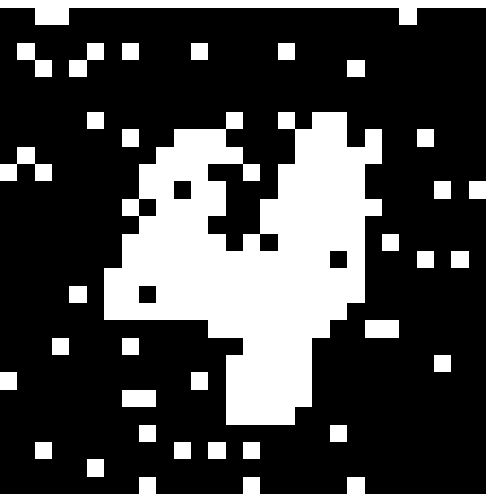}
	}
	\subfloat{
		\label{4_784}
		\includegraphics[width=2 cm]{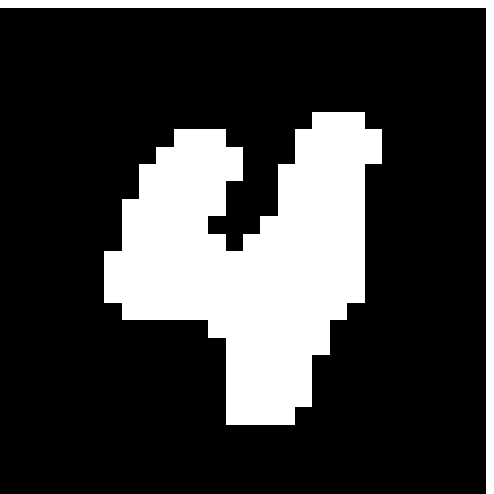}
	}	
	
	\caption{Denoising of Patterns (from $0$ to $4$) with $30\%$ Noise  \\ at epochs $1, 300, 500, 600, 784$} 
	\label{fig:denoisedPatterns1}
\end{figure}

\begin{figure}[!htb] 
	\centering		
	\subfloat{
		\label{5_0}
		\includegraphics[width=2 cm]{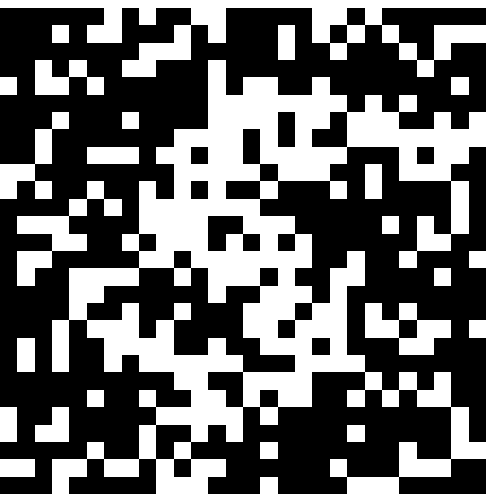}
	}
	\subfloat{
		\label{5_300}
		\includegraphics[width=2 cm]{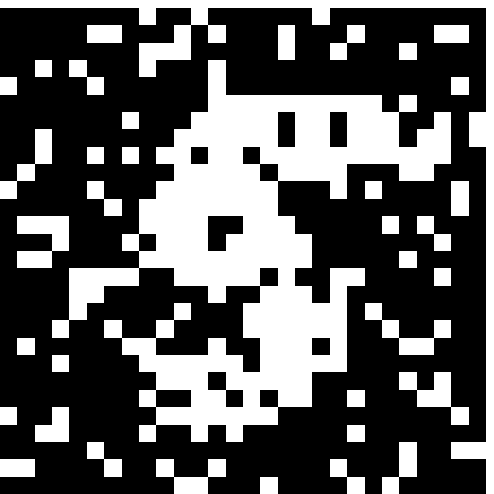}
	}
	\subfloat{
		\label{5_500}
		\includegraphics[width=2 cm]{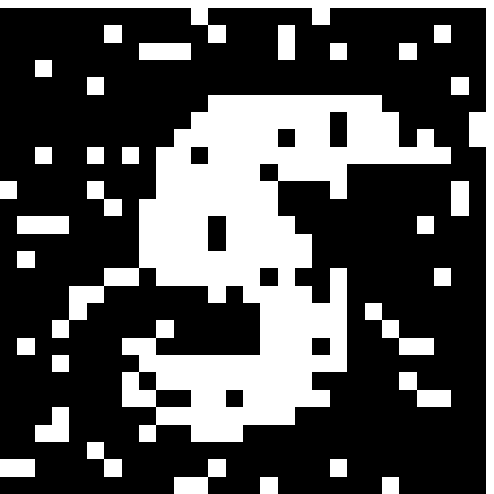}
	}
	\subfloat{
		\label{5_600}
		\includegraphics[width=2 cm]{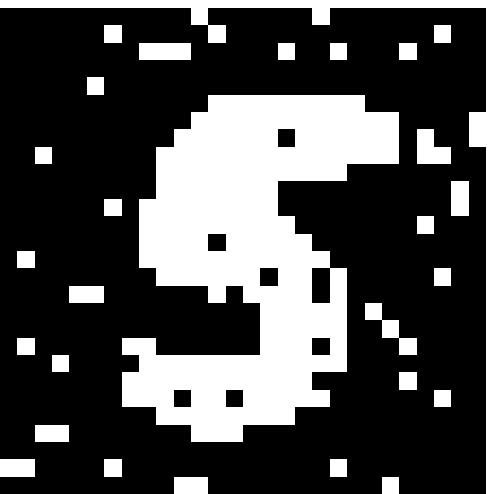}
	}
	\subfloat{
		\label{5_784}
		\includegraphics[width=2 cm]{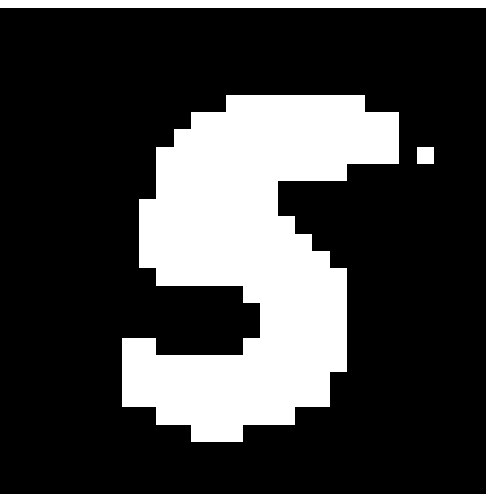}
	} 
	\\
	\subfloat{
		\label{6_0}
		\includegraphics[width=2 cm]{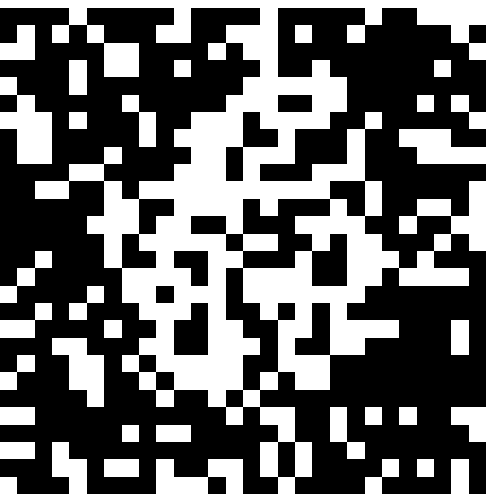}
	}
	\subfloat{
		\label{6_300}
		\includegraphics[width=2 cm]{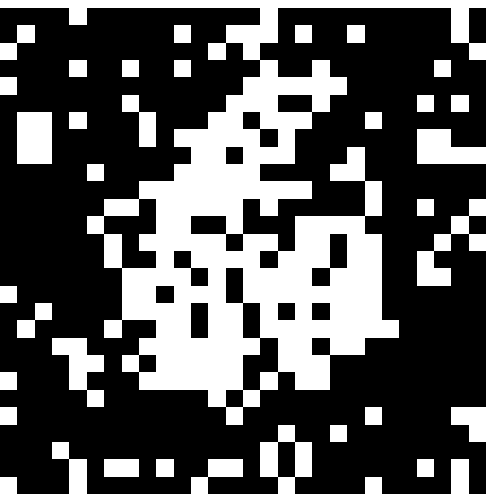}
	}
	\subfloat{
		\label{6_500}
		\includegraphics[width=2 cm]{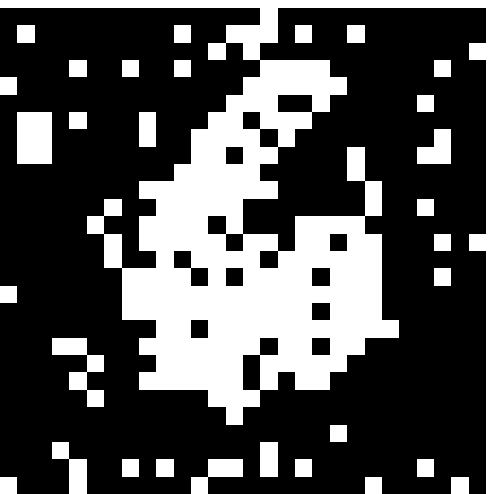}
	}
	\subfloat{
		\label{6_600}
		\includegraphics[width=2 cm]{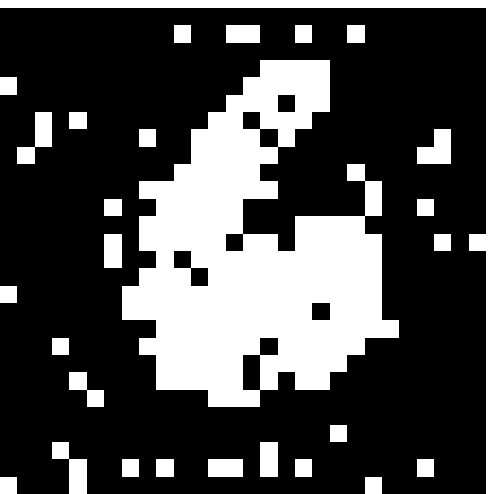}
	}
	\subfloat{
		\label{6_784}
		\includegraphics[width=2 cm]{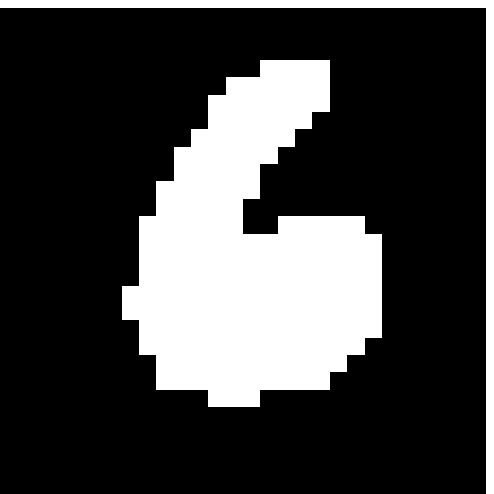}
	} 
	\\
	\subfloat{
		\label{7_0}
		\includegraphics[width=2 cm]{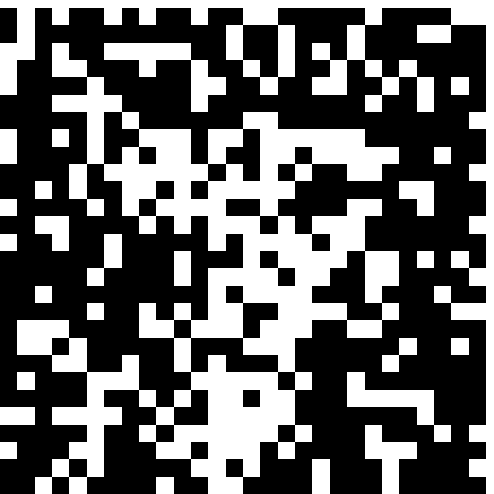}
	}
	\subfloat{
		\label{7_300}
		\includegraphics[width=2 cm]{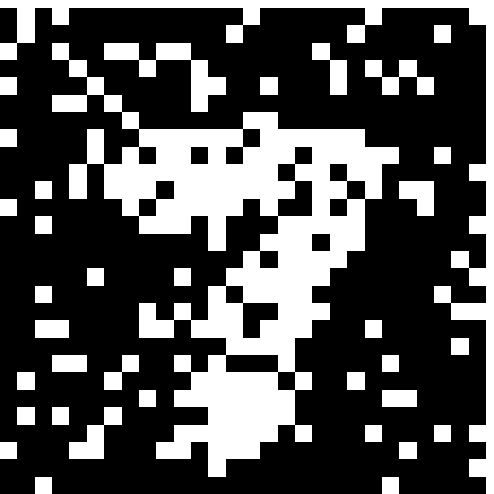}
	}
	\subfloat{
		\label{7_500}
		\includegraphics[width=2 cm]{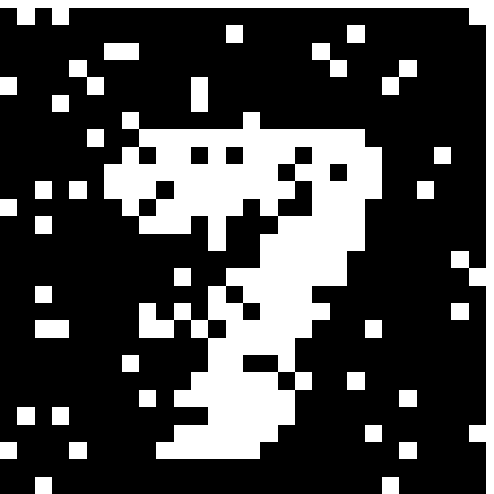}
	}
	\subfloat{
		\label{7_600}
		\includegraphics[width=2 cm]{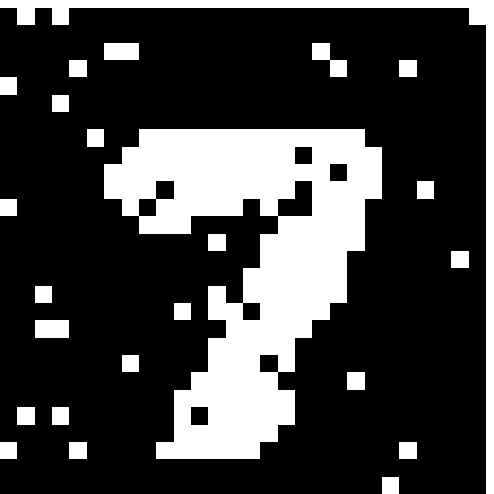}
	}
	\subfloat{
		\label{7_784}
		\includegraphics[width=2 cm]{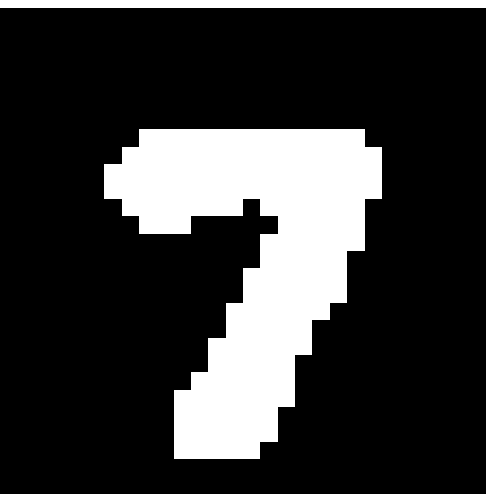}
	} 
	\\
	\subfloat{
		\label{8_0}
		\includegraphics[width=2 cm]{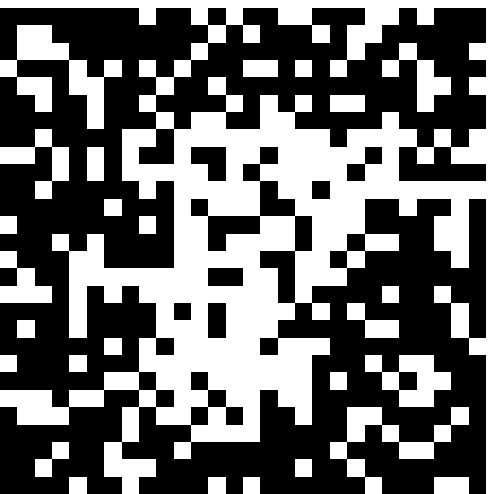}
	}
	\subfloat{
		\label{8_300}
		\includegraphics[width=2 cm]{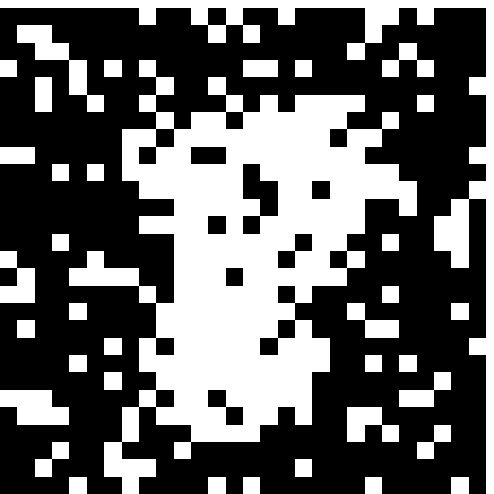}
	}
	\subfloat{
		\label{8_500}
		\includegraphics[width=2 cm]{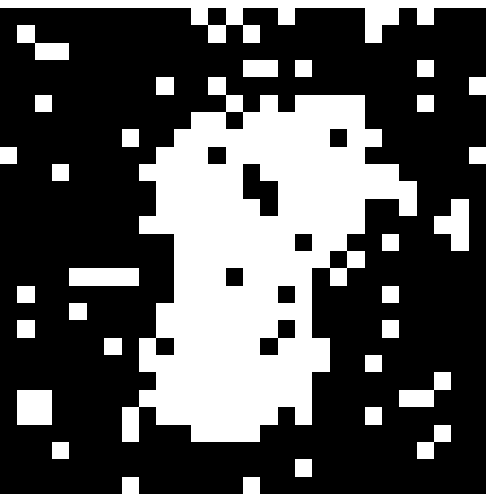}
	}
	\subfloat{
		\label{8_600}
		\includegraphics[width=2 cm]{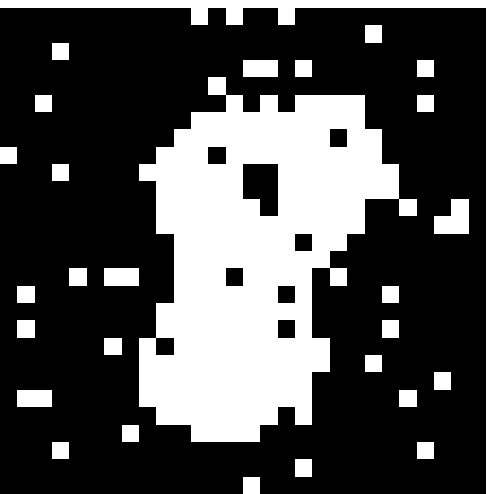}
	}
	\subfloat{
		\label{8_784}
		\includegraphics[width=2 cm]{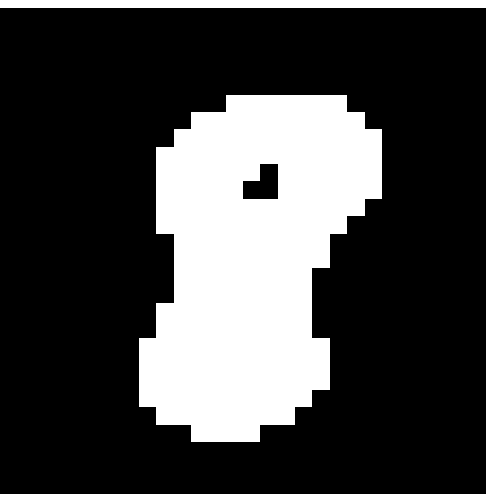}
	} 
	\\
	\subfloat{
		\label{9_0}
		\includegraphics[width=2 cm]{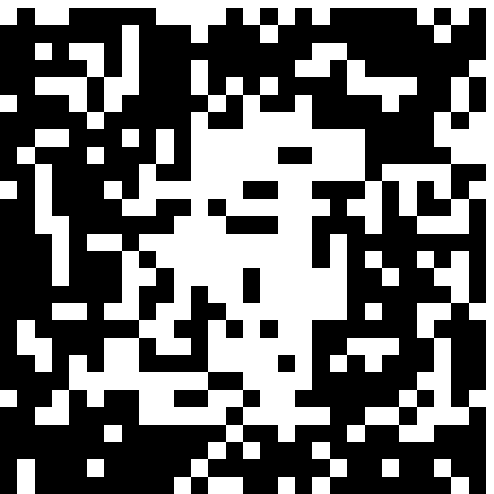}
	}
	\subfloat{
		\label{9_300}
		\includegraphics[width=2 cm]{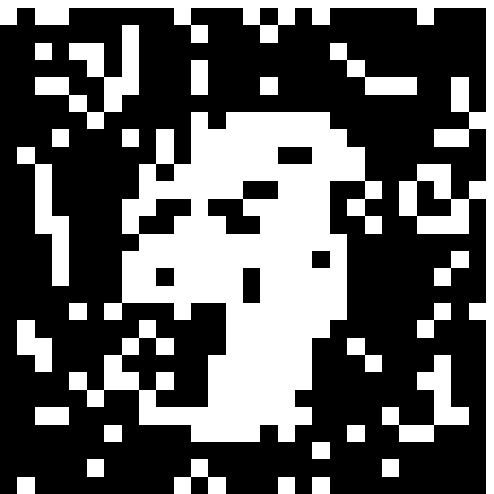}
	}
	\subfloat{
		\label{9_500}
		\includegraphics[width=2 cm]{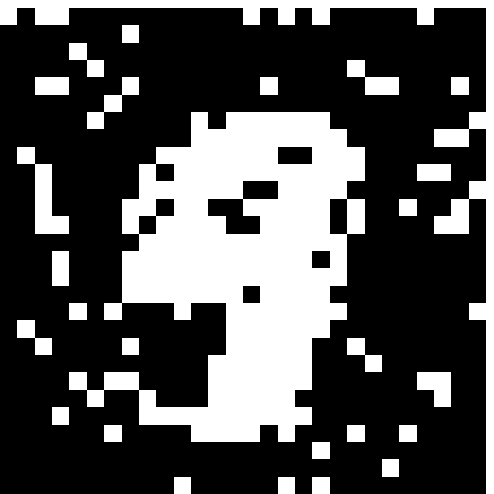}
	}
	\subfloat{
		\label{9_600}
		\includegraphics[width=2 cm]{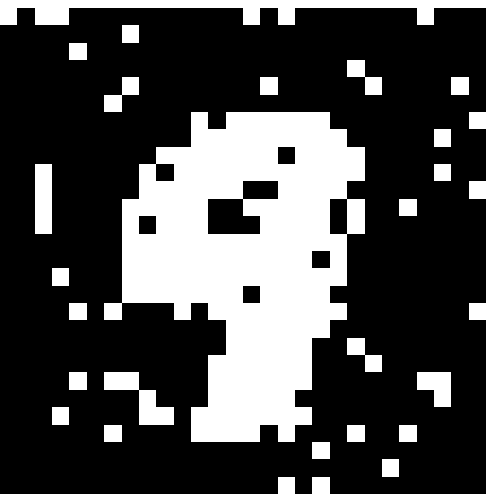}
	}
	\subfloat{
		\label{9_784}
		\includegraphics[width=2 cm]{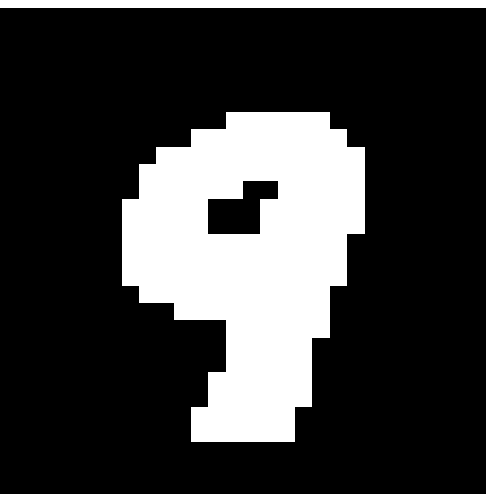}
	}     
	\caption{Denoising of Patterns  (from $5$ to $9$) with $30\%$ Noise \\ at epochs $1, 300, 500, 600, 784$} 
	\label{fig:denoisedPatterns2}
\end{figure}

\begin{figure}[!htb]
	\includegraphics[angle=-90, width=8cm]{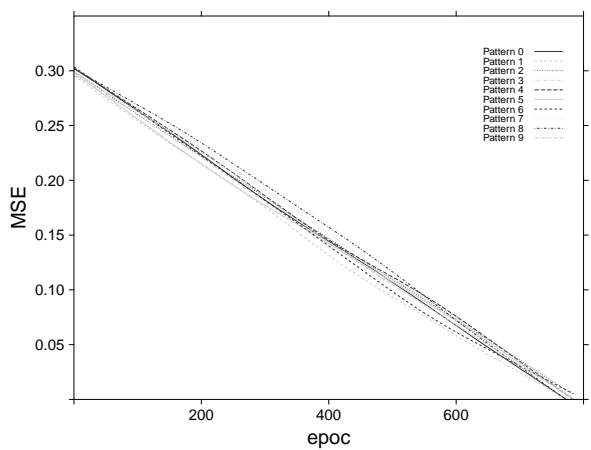}
	\centering
	\caption{MSEs for Typical Samples (from $0$ to $9$) of $30\%$ Noise during Denoising Epocs} 
	\label{fig:mse}
\end{figure}

	\begin{table}[!htb] 
		\centering
		\begin{tabular}{cc|c@{\hskip 0.2in} c@{\hskip 0.2in}|c}
			\multicolumn{4}{c}{\hspace{4ex}Recalled ?} \\
			\multicolumn{2}{c|}{} & yes & no & Acc. (\%)\\
			\hline
			\multirow{10}{*}{\begin{sideways} Patterns~ \end{sideways}} 
			
			& $0$ & $100$ & $0$ & $100$\\
			
			& $1$ & $100$ & $0$ & $100$\\
			
			& $2$ & $99$ & $1$ & $99$\\
			
			& $3$ & $100$ & $0$ & $100$\\
			
			& $4$ & $100$ & $0$ & $100$\\
			
			& $5$ & $100$ & $0$ & $100$\\
			
			& $6$ & $100$ & $0$ & $100$\\
			
			& $7$ & $100$ & $0$ & $100$\\
			
			& $8$ & $87$ & $13$ & $87$\\
			
			& $9$ & $99$ & $1$ & $99$\\
			\hline 
			\multicolumn{2}{c}{} &
			\multicolumn{2}{c}{Ave. Acc. (\%) :}  & $98.5$ \\
			\hline 
		\end{tabular}
		\captionof{table}{Recall Accuracy at $30\%$ Noise}
		\label{tab:tab2}
	\end{table}

Our model recalls the stored pattern given the noisy pattern by computing the weighted average of local clique potentials; which are defined using the hybridization reactions. The average recall accuracies for noisy patterns at different noise percentages are depicted in figure~\ref{fig:accVsNoise}. The average recall accuracies are high ($>98.5\%$) up to $30\%$ of noise. The accuracies drop owing to heavy randomness over $30\%$ of noise. The average recall accuracies at $30\%$ of noise for each of the patterns are shown in table~\ref{tab:tab2}. 

On a successful recall, we denoise the noisy pattern iteratively  involving hybridization reactions. We present both qualitative (refer figures~\ref{fig:denoisedPatterns1} and~\ref{fig:denoisedPatterns2}) and quantitative results of denoising (refer figure~\ref{fig:accVsNoise}). The denoising of typical patterns (digits from $0$ to $9$) with $30\%$ of noise at epochs $1, 300, 500, 600, 784$ are shown in figures~\ref{fig:denoisedPatterns1} and~\ref{fig:denoisedPatterns2}. We compute Mean Squared Error (MSE) at each epoch of denoising of typical patterns and are shown in figure~\ref{fig:mse}. We notice linear reconstruction of all the patterns with our proposed denoising molecular algorithm. The averaged MSEs of all the patterns at different noise percentages are shown in figure~\ref{fig:accVsNoise}. MSEs are low ($<0.014$) up to $30\%$ of noise.
\section{Conclusion}
\label{sec:conc}
\vspace*{-0.4cm}
We demonstrate that associative memory can be realized on a molecular level, involving only local features using Pairwise Markov Random Field (PMRF) models. We apply DNA based bio-molecular operations with PMRF models for extracting, storing, learning, recalling and denoising of information. The results show that our proposed molecular simulation of associative memory denoises information with low MSE ($<0.014$) up to $30\%$ of noise. Our molecular computation model, like the human brain, is able to recall and reconstruct (denoise) the patterns when noisy patterns are provided.
 
\vspace*{-0.5cm}
\section{Acknowledgments}\vspace*{-0.3cm}
This work was partly supported by  Samsung Research Funding Center of Samsung Electronics (SRFC-IT1401- 12), the Institute for Information \& Communications Technology Promotion (2015-0-00310-SW.StarLab, 2017-0-01772-VTT, 2018-0-00622-RMI, 2019-0-01367-BabyMind) and Korea Institute for Advancement Technology (P0006720-GENKO) grant funded by the Korea government. ICT at Seoul National University provided research facilities for the study.

\bibliographystyle{unsrt}
\vspace*{-0.4cm}
\bibliography{dnaMem} 

\end{document}